\documentstyle[sprocl,epsfig]{article}

\def\be{\begin{equation}}
\def\ee{\end{equation}}
\def\bea{\begin{eqnarray}}
\def\eea{\end{eqnarray}}
\def\nn{\nonumber}
\newcommand{\lsim}{\raisebox{-3pt}{$\,\stackrel{\textstyle <}{\sim}\,$}}

\begin{document}

\title{GENERALIZED PARTON DISTRIBUTIONS AT LARGE MOMENTUM TRANSFER \footnote{
Talk presented at the 9th Intern.\ Workshop on Deep Inelastic
Scattering, Bologna(2001).}} 

\author{P. KROLL}

\address{Fachbereich Physik, Universit\"at Wuppertal, D-42097 Wuppertal, 
Germany\\E-mail: kroll@theorie.physik.uni-wuppertal.de}

\maketitle\abstracts{The role of generalized parton distributions 
in wide-angle exclusive reactions will be discussed. In contrast to
deep virtual exclusive reactions the wide angle processes offer the
possibility of investigating the generalized parton distributions at
large momentum transfer.} 

\section{Introduction}
\begin{figure}[hb]
\parbox{\textwidth}
{\begin{center}
   \psfig{figure=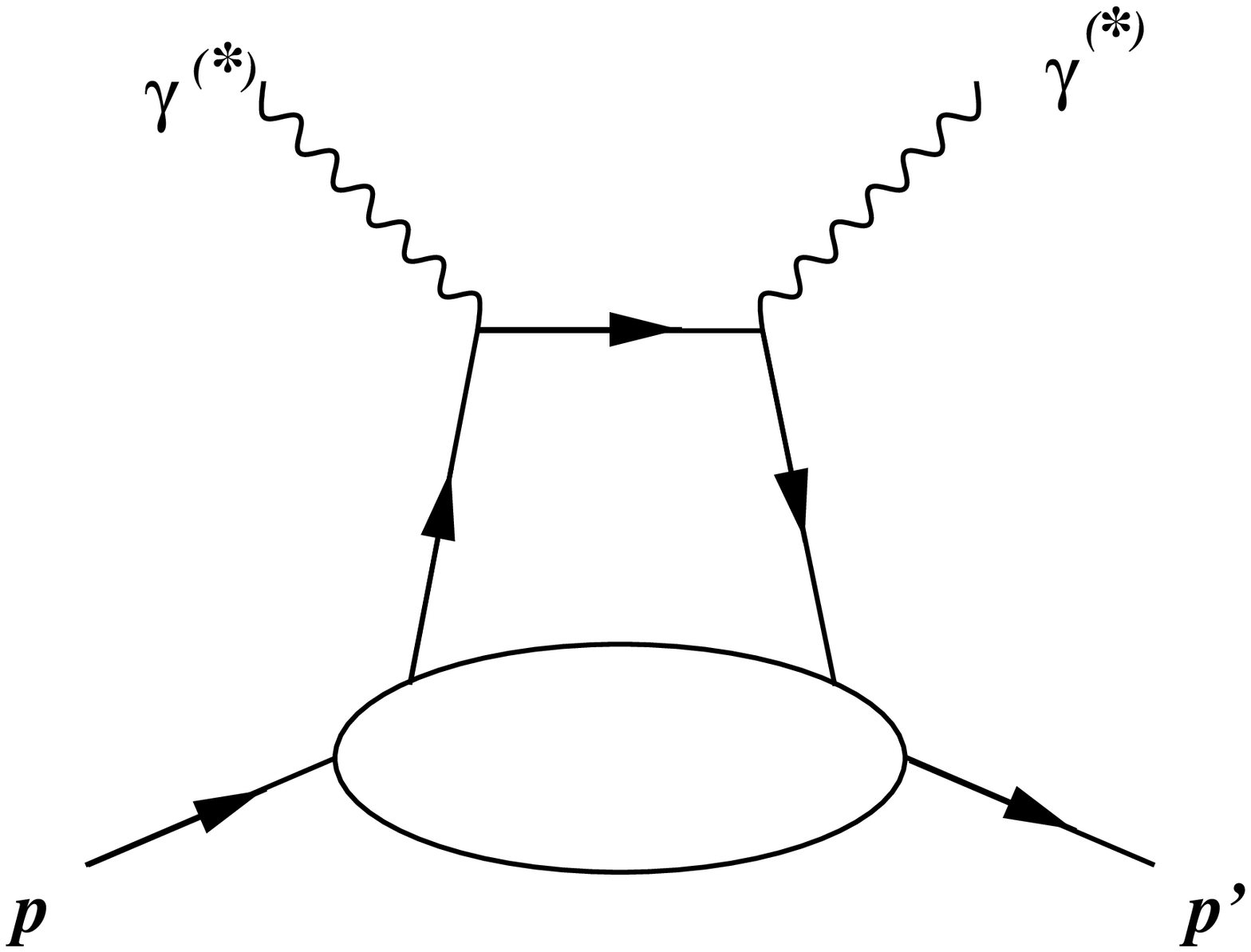,%
          bbllx=45pt,bblly=230pt,bburx=545pt,bbury=610pt,%
           width=4.0cm,clip=} 
\hspace{1cm}
   \psfig{figure=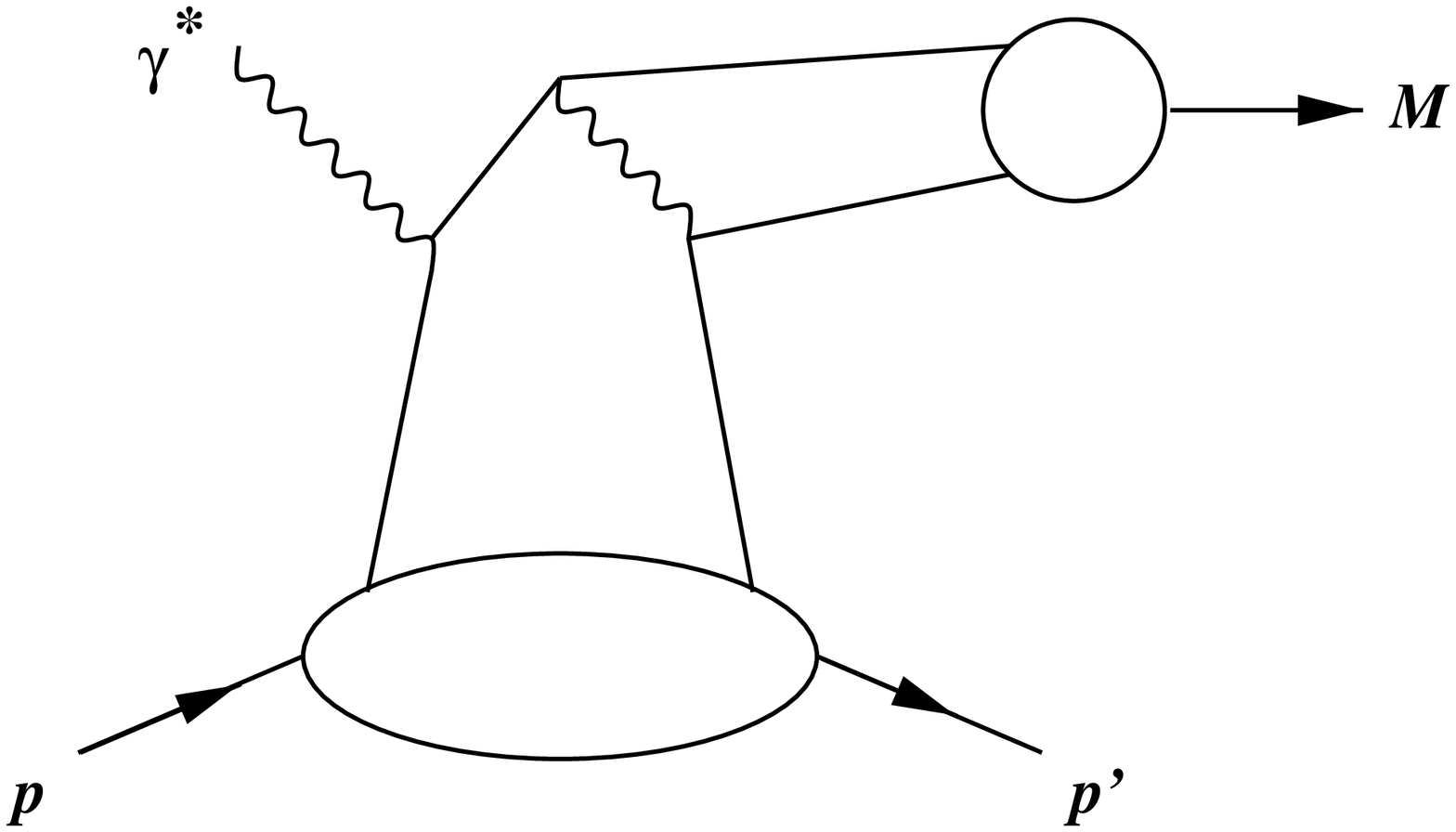,%
          bbllx=15pt,bblly=265pt,bburx=570pt,bbury=580pt,%
           width=4.5cm,clip=}
\end{center}}
\caption{The handbag diagram for Compton scattering (left) and for
meson electroproduction (right).}
\label{fig:handbag}
\end{figure}

Recently, deep virtual exclusive reactions such as Compton scattering
or hard meson electroproduction at large photon virtualities, $Q^2$,
and small momentum transfer, $-t$, attracted much interest. One can
show \cite{rad97,ji98,rad96} that the relevant amplitudes factorize in
parton-level amplitudes and process independent soft proton matrix
elements which represent generalized parton distributions (GPDs), see 
Fig.\ \ref{fig:handbag}. One may also consider Compton scattering or 
meson electroproduction in the wide-angle region where $-t$ (and $-u$) 
is large but $Q^2$ small. These processes are controlled by the
handbag diagram shown in Fig.\ \ref{fig:handbag} too and their amplitudes 
also factorize in parton-level amplitudes and GPDs \cite{DFJK1,hanwen00}. 
The analyses of deep virtual and wide-angle reactions provide
complementary informations on GPDs, namely either on their small $-t$
or on their large $-t$ behaviour, respectively. In this talk I am
going to report on studies of the wide-angle exclusive reactions.  
\section{GPDs and soft physics}
\begin{figure}[t]
\parbox{\textwidth}{\begin{center}  
   \psfig{figure=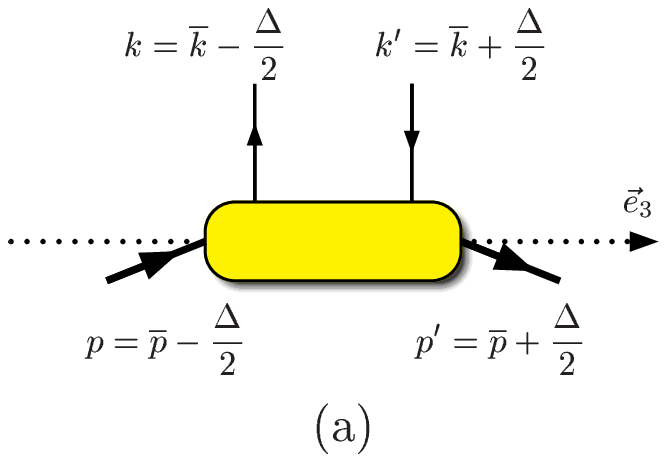,%
          bbllx=190pt,bblly=460pt,bburx=380pt,bbury=580pt,%
           width=4.5cm,clip=}\hspace{1.5cm}
   \psfig{file=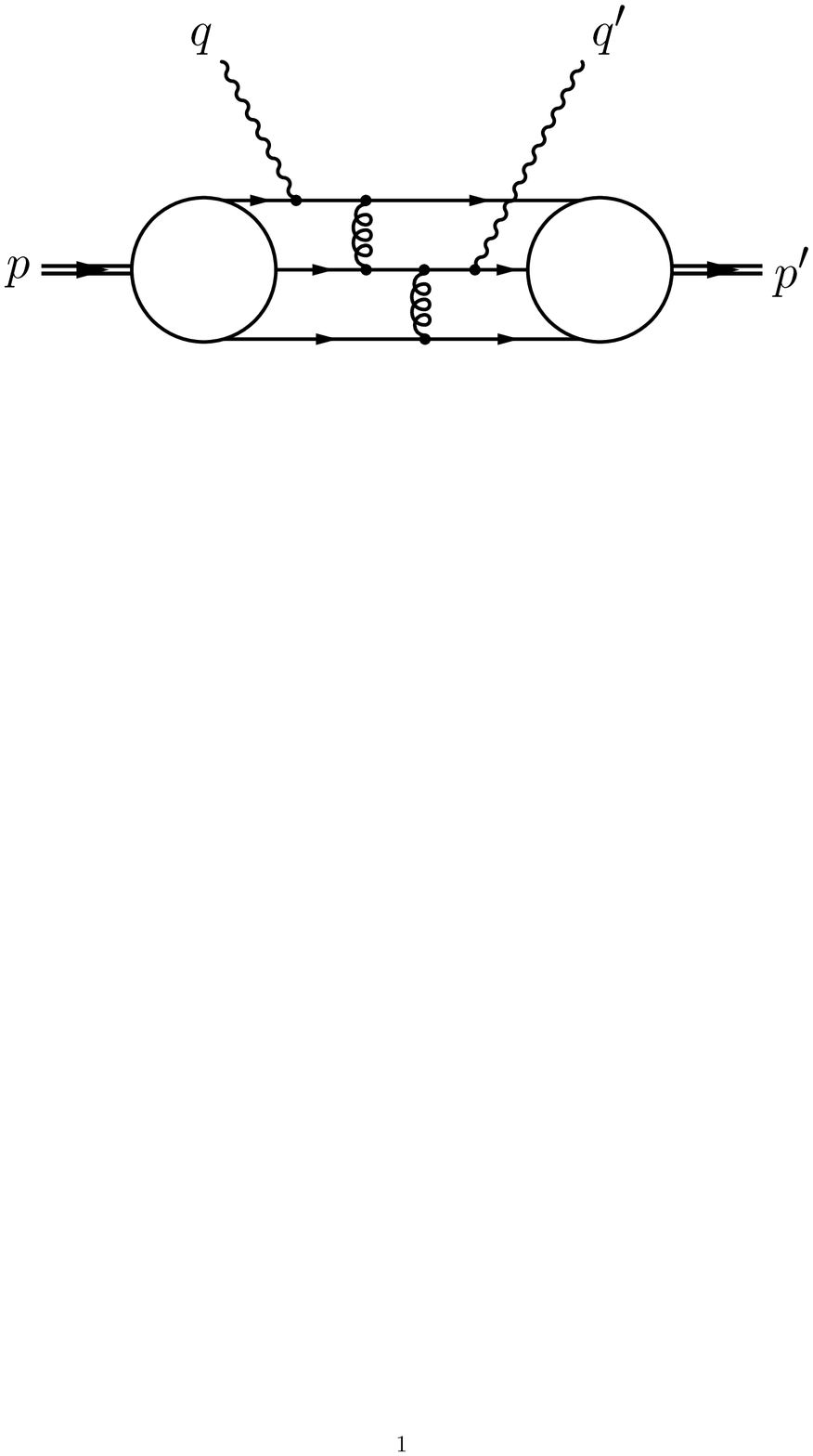,%
          bbllx=90pt,bblly=585pt,bburx=510pt,bbury=790pt,%
           width=5.0cm,clip=}
\vspace{-0.3cm}
\end{center}}
\caption{The kinematics for GPDs (left) and a typical Feynman graph
for Compton scattering at asymptotically large $-t$.}
\label{fig:kinematics}
\end{figure}

The GPDs $H^q(\bar{x},\xi;t)$ ($\widetilde{H}^q$) and $E^q(\bar{x},\xi;t)$
($\widetilde{E}^q$) for a quark of flavour $q$ are defined as the
Fourier transform of a bilocal product of field operators and
$\gamma^+$ ($\gamma^+\gamma_5$) \cite{rad97,ji97}. The fractions of 
light-cone plus momentum components appearing as arguments of the
GPDs, are given by (see Fig.\ \ref{fig:kinematics})
\begin{equation}
\xi=\frac{(p-p^\prime)^+}{(p+p^\prime)^+}\,, \qquad 
\bar{x}=\frac{(k+k^\prime)^+}{(p+p^\prime)^+}\,.
\label{eq:fractions}
\end{equation}
The GPDs describe the emission of a quark with momentum fraction 
$(\bar{x}+\xi)/(1+\xi)$ from the proton and the reabsorption of a
quark with momentum fraction $(\bar{x}-\xi)/(1-\xi)$. Reduction 
formulas relate the GPDs to the ordinary parton distributions:
\be
\label{eq:red1}
H^q(\bar{x},0;0) = q(\bar{x}), \phantom{x} \qquad  
\widetilde{H}^q(\bar{x},0;0)\; = \;\Delta q(\bar{x})\,.
\ee
Integrating the GPDs over $\bar{x}$, one obtains the
contributions of flavour $q$ quarks to the proton form factors,
e.g.,
\begin{equation}
\label{eq:red2}
F_1^q(t) = \langle \bar{x}^0 \rangle = \int_{-1}^1 d\bar{x}\, 
                                 H^q(\bar{x},\xi;t) \,.
\end{equation}
Multiplying $F^q_1$ by the appropriate electric charges and summing
over all flavours, one obtains the full Dirac form factor. The 
$\langle \bar{x}^0 \rangle$-moments $E^q$ provides contributions to 
the Pauli form factor $F_2$.

As is well-known, for asymptotically large momentum transfer
wide-angle exclusive scattering is controlled by perturbative
contributions where all partons the proton is made off, participate in 
the hard scattering, see Fig.\ \ref{fig:kinematics}. There is evidence
that the perturbative contributions to the proton form factor
\cite{bol96,ber95} and to Compton scattering \cite{brooks} are way
below experiment provided plausible proton wave functions are used.
The soft handbag contributions depicted in Fig.\ \ref{fig:handbag}, 
seem to dominate at finite values of momentum transfer
\cite{DFJK1,hanwen00,bol96,rad98a} despite of the fact that they
formally represent power corrections to the perturbative ones.
The soft contributions are defined \cite{DFJK1} through the assumption
that the (soft) light-cone wave functions (LCWFs) occuring in the Fock
decomposition of the proton, are dominated by parton virtualities 
$k_i^2, k'{}^2_i \lsim \Lambda^2$ and by intrinsic transverse momenta 
that satisfy $\tilde{k}^2_{\perp i}/\tilde{x}_i,\: 
\hat{k}'{}^2_{\perp i}/\hat{x}'_i \lsim \Lambda^2$ where $\Lambda$ 
represents a typical hadronic scale of order 1 GeV. Using a symmetric
frame in which $\Delta^+=\xi=0$, one can show that the active partons,
i.e. those to which the photons couple (see Fig.\ \ref{fig:handbag}), 
are approximately on-shell, collinear with their parent hadrons and carry
momentum fractions close to unity. Thus, the physical situation is
that of a hard photon-parton scattering and a soft emission and
re-absorption of partons by the protons. In this situation the
hadronic amplitudes factorize into parton-level amplitudes (either 
$\gamma^* q\to \gamma q$ or $\gamma^* q \to M q$) and
$1/\bar{x}$-moments of $\xi=0$ GPDs. The proton helicity conserving
Compton amplitudes read \cite{DFJK1}
\bea
{\cal M}_{\mu'+,\,\mu +}(s,t) \,&=&\, \;2\pi\alpha_{\rm em} \left[\,{\cal
    H}_{\mu'+,\,\mu+}(s,t)\,(R_V(t) + R_A(t)) \,\right. \nn\\
    &&\left. \hspace{1cm}+ \, {\cal H}_{\mu'-,\,\mu-}(s,t)\, 
                                     (R_V(t) - R_A(t)) \,\right ]\,.
\label{final}
\eea 
$\mu$ ($\mu'$) denotes the helicity of the incoming (outgoing) photon.
The ${\cal H}_{\mu'\nu,\mu\nu}$ denote the familiar amplitudes for 
Compton scattering off massless quarks (with helicity $\nu$).
The soft form factors, $R_V$ and $R_A$, for active quarks of
flavour $q$, are given by
\begin{equation}
R_V^{\,q}(t) = \int_{-1}^1
\frac{d\bar x}{\bar x}\;H^q(\bar x, 0;t)\,, \qquad
R_A^{\,q}(t) = \int_{-1}^1 
\frac{d\bar x}{\bar x}\;
\mbox{sign}(\bar x)\,
\widetilde{H}^q(\bar x, 0;t)\,.
\label{eq:cmoments}
\end{equation}
The full form factors are specific to the process under consideration. 
All charged partons contribute to Compton scattering, (e.g.\ $R_V(t) =
\sum e_q^2 R_V^q(t)$) while in electroproduction the meson selects its
valence quarks from the proton. The amplitudes for wide-angle photo-
and electroproduction of mesons have a representation similar to
(\ref{final}) \cite{hanwen00}. 

In principle there is third form factor, $R_T$, which
represents a $1/\bar{x}$-moment of the GPD $E^q$. Note that $E^q$
involves parton orbital angular momentum in an essential way. On the
basis of the overlap representation of the GPDs \cite{DFJK3} one can
argue that $R_T/R_V\simeq F_2/F_1$ at large $-t$. Since the SLAC data
on $F_2$ \cite{and94} indicate a behaviour $F_2/F_1 \simeq -m^2/t$ and
given that the evaluation of the handbag diagram is only accurate up to
corrections in $\Lambda^2/t$, $R_T$ and, hence, proton helicity flip
have been neglected in \cite{DFJK1,hanwen00} for consistency. 
The new Jlab data on $F_2$ \cite{jon99}, however, seems to be
compatible with a $m/\sqrt{-t}$ behaviour of the ratio of form factors
rather than $-m^2/t$. Provided this result will be confirmed, $R_T$ is
to be taken into account. Note that a behaviour $\propto m/\sqrt{-t}$
for the ratio of form factors appears quite natural in the overlap
representation (see also Ralston's talk \cite{ral01}).  
\section{Predictions for Compton scattering}
In order to predict wide-angle exclusive reactions a model for the
GPDs is required. It has been shown on the basis of light-cone
quantisation \cite{DFJK1,DFJK3} that the GPDs possess a representation
in terms of LCWF overlaps. It allows to construct a simple model for
the GPDs by parameterizing the transverse momentum dependence of the
LCWFs as 
\begin{equation}
\Psi_{N} \propto \exp{\left[-a_N^2 \sum^N_{i=1}
                                  k^2_{\perp i}/x_i\right]}\,,   
\label{eq:gaussian}
\end{equation}
which is in line with the central assumption of the soft physics approach 
of restricted $k^2_{\perp i}/x_i$, necessary to achieve the
factorization of the amplitudes into soft and hard parts. Without explicit
specification of the $x$-dependence of the LCWFs one can then evaluate
the $\xi=0$ GPDs from the overlap representation if a common
transverse size parameter $a=a_N$ is used. One obtains
\begin{equation}
H^q(\bar{x}, 0;t)=\, \exp{\left[a^2 t\,
               \frac{1-\bar{x}}{2\bar{x}}\right]}\, q(\bar{x})\,,
\;
\widetilde{H}^q(\bar{x}, 0;t)=\, \exp{\left[a^2 t\,
               \frac{1-\bar{x}}{2\bar{x}}\right]}\, \Delta q(\bar{x})
\,.
\label{xi0}
\end{equation}
Taking the parton distributions from one of the current analyses of
deep inelastic lepton-nucleon scattering and using a value of $\simeq 1$
GeV$^{-1}$ for the transverse size parameter, one finds reasonable 
results for the proton and neutron form factors. Improvements can be
obtained by treating the lowest Fock states explicitly with specified
$x$-dependencies \cite{DFJK1,bol96}.

\begin{figure}
\parbox{\textwidth}{\begin{center}
   \psfig{file=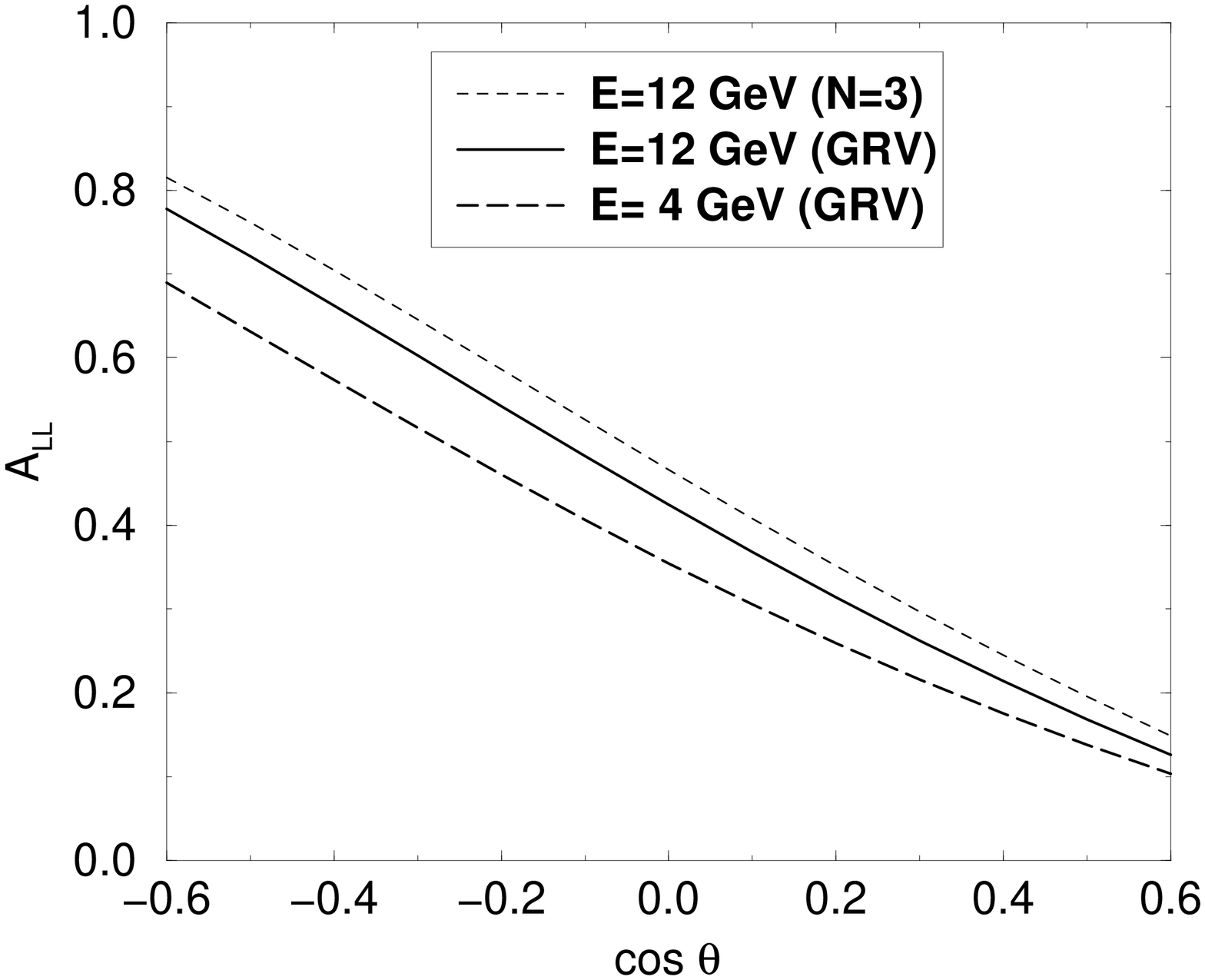,%
          width=4.0cm, angle=-90} \hspace{0.5cm}
   \psfig{file=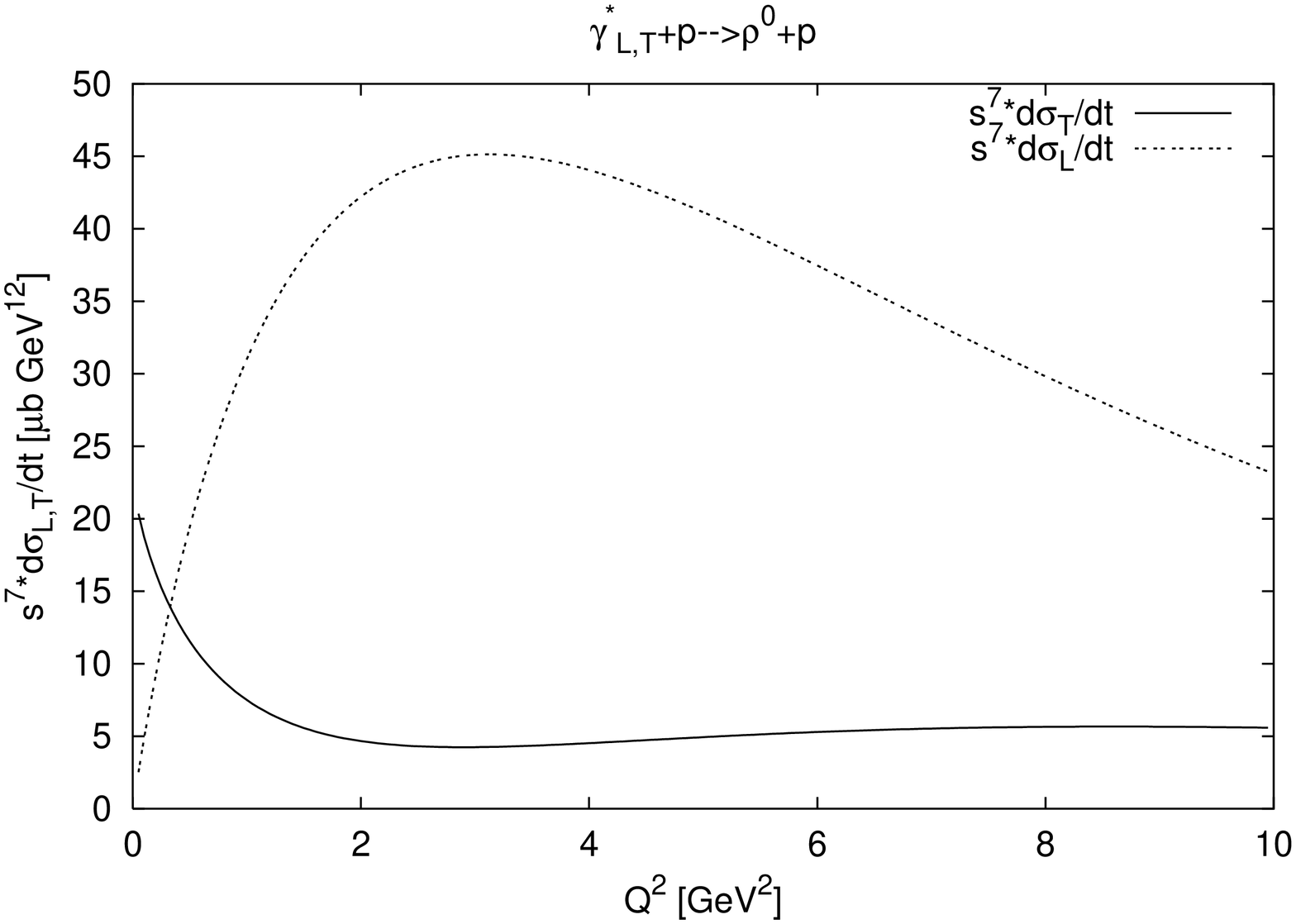, bbllx=95pt,bblly=55pt,bburx=560pt,bbury=750pt,%
                 width=4.0cm,angle=-90, clip=} 
\vspace{-0.3cm}
\end{center}}
\caption{Predictions from the soft physics approach for the helicity
         correlation $A_{\rm LL}$ (left)  and for the transverse and 
         longitudinal electroproduction cross sections of longitudinally 
         polarised $\rho^0$ mesons at $s=40\, {\rm GeV}^2$ and a cm.\ 
         scattering angle of $90^{\circ}$ (left).}
\label{fig:all}
\end{figure}

The amplitude (\ref{final}) leads to the following Compton cross
section 
\begin{equation}
\frac{{d} \sigma}{{d} t} = \frac{{d} \hat{\sigma}}{{d} t}
                       \left [\, \frac{1}{2} (R_V^2(t) + R_A^2(t))
        -\, \frac{us}{s^2+u^2}\, (R_V^2(t)-R_A^2(t)) \,\right] \,,
\label{eq:rcs-cs}
\end{equation}
where $d\hat{\sigma}/dt$ is the Klein-Nishina cross section for
Compton scattering off point-like fermions. The predictions for this
cross section are in fair agreement with experiment. The initial state 
helicity correlation approximately reads 
\begin{equation}
A_{\rm LL}\, \simeq \frac{s^2 - u^2}{s^2 + u^2} \frac{ R_A(t)}{ R_V(t)}\,.
\label{eq:all}
\end{equation}
Precise data on both observables would permit an extraction of the form
factors, $R_V$ and $R_A$, quite similar to that of the Dirac and Pauli 
form factors and, hence, would allow a model independent test of the
soft physics approach. Predictions for $A_{LL}$ are shown in Fig.\
\ref{fig:all}. It is to be stressed that $R_T$ has been neglected in 
(\ref{eq:rcs-cs}) and (\ref{eq:all}); its inclusion will lead to
corrections of about $10\%$. The corresponding correlation between the 
photon helicity and the sideway polarization of the proton (normal to 
its momentum but in the scattering plane) is particular sensitive to
$R_T$, namely $A_{LS}/A_{LL} \propto R_T/R_V$. There are, however, 
substantial corrections to this result at moderately large values of 
$-t$.   

As an example of wide-angle electroproduction of mesons, the $\rho^0$ 
cross sections for longitudinally and transversally polarised photons 
are shown in Fig.\ \ref{fig:all}. The longitudinal cross section 
dominates except for $Q^2 \lsim 1 {\rm GeV}^2$. Contrary to a
statement to be found in the literature occasionally the soft physics 
approach leads to a $s^7$-scaling of the cross sections provided $t/s$ 
and $Q^2/s$ are kept fixed and to the extent that the form factors 
$R^M_{V(A)}$ behave $\propto 1/t^2$. Unfortunately, there is no 
wide-angle electroproduction data available as yet to compare
with. For more details and predictions it is refered to
\cite{hanwen00}. Predictions for the $ep\to ep\gamma$ 
cross sections have also been given in Ref.\ \cite{DFJK1}. 
   
\section{Summary}
The GPDs are new tools for the description of soft hadronic matrix
elements. They are central elements which connect many different 
inclusive and exclusive processes: polarised and unpolarised parton 
distributions are the $\xi=t=0$ limits of GPDs, electromagnetic and 
Compton form factors represent moments of the GPDs, deeply virtual
Compton scattering and hard meson electroproduction are controlled by
them. A particularly interesting aspect is touched in exclusive
reactions such as proton form factors and wide-angle Compton scattering. 
Their analysis by means of GPDs implies the calculation of soft physics
contributions to these processes in which only one of the quarks is
considered  as active while the others act as spectators. The soft
contributions formally represent power corrections to the asymptotically 
leading perturbative contributions in which all quarks participate in 
the subprocess. There is evidence that for momentum transfers around 
10 GeV$^2$ the soft contribution dominates over the perturbative one. 
However, a severe confrontation of this approach with accurate data on
wide-angle Compton scattering and electroproduction of mesons is pending.
\section*{Acknowledgments}
This work has been supported by the TMR network
HPRN-CT-2000-00130.

\end{document}